\documentstyle[aps,preprint]{revtex}
\begin{document}
\draft{}
\title{ISOBAR ELECTROPRODUCTION AS A BACKGROUND FROM INTERACTION OF BEAMS
WITH RESIDUAL GAS AT $\phi$-FACTORIES.}
\author{M.N. Achasov\thanks{G.I. Budker Institute for Nuclear Physics,
\ \ \ Novosibirsk-90,\ \ \ 630090,\ \ \ Russia},
N.N. Achasov, V.B. Golubev $ ^\ast$ and S.I. Serednyakov $^\ast$ \\
Laboratory of Theoretical Physics \\
S.L. Sobolev Institute for Mathematics \\
Novosibirsk-90,\ \ \  630090,\ \ \ Russia
\thanks{E-mail: achasov@math.nsc.ru}}
\date{\today}
\maketitle
\begin{abstract}
It is shown that when beams interact  with a residual gas at $\phi$-factories the reaction of the electroproduction of the 
$\Delta (1232)$ isobar proceeds vigorously. The isobar decay gives 
$\sim 10^7$ pions during an effective year of $10^7$ s per meter of a 
residual gas. These pions are emitted largely across the beam axis and
have a resonance energy distribution with a peak nearby 265 MeV of a width 
close to 120 MeV in the isobar rest system. There are presented formulae for
the distributions of the four-momentum transfer square, the angles, the
energies and the momentum of the decay products, that is all required 
for the simulation of the process under consideration.
\end{abstract}


The aim of $\phi$-factories (DA$\Phi$NE in Frascati, which is to be launched
in 1997  \cite{pdg-96}, and another one, which is builing up in Novosibirsk
\cite{pdg-96,ans}) is to carry out the precision measurements of the
most important physical values, $\epsilon ^\prime /\epsilon $ first
\cite{Maiani-95}.

To realize this  program  it  needs  much to study facilities, all most
essential backgrounds. A source of such backgrounds is an interaction of
beams with a residual gas.

In the paper  we show that the  cross section of  the electroproduction of
the isobar $\Delta (1232)$  with $I(J^P)=
\frac{3}{2}\left (\frac{3}{2}^+\right )$\cite{pdg-96} at a nucleon for the
electron (positron) energy 509.5 MeV (the $\phi$-factories energy) is equal
to 3 $\mu$b. It results in the production of $\sim 10^7$ isobars during an 
effective year of $10^7$ s per meter of a residual gas  at  the total 
electric current of beams $ 1.3 \to 5.2$ A and  the residual gas pressure
$\sim 1 \mbox{nTorr}$ (as projected at DA$\Phi$NE \cite{leefranzini-95}).

From the isobar decay $(\Delta(1232)\to\pi~N~~\mbox{and}~~\gamma~N)$ there
are produced the $\pi$-mesons, having the resonance energy distribution with
the peak nearby 265 MeV of a width close to 120 MeV, emitted largely
across the beam axis, and the photons, having the resonance energy 
distribution with the peak nearby 257 MeV of a width close to 120 MeV, 
emitted more isotropically.

For the process under discussion it is characteristic that protons and 
neutrons, emitted largely across the beam axis from isobar decay, have the 
very narrow energy resonance distribution (the width is 30 MeV) nearby 970 
MeV.

We present formulae for distributions of the four-momentum transfer
square of the electron (positron), for distributions of the angles of decay 
pions (nucleons) and decay photons (nucleons) in the isobar rest system, for 
distributions of the pion, nucleon and photon energies in the isobar rest 
system and also for distributions of the pion (nucleon) momentum in the 
isobar rest system, that is all required for the simulation and the
separation of the background under discussion.

The phenomenological Lagrangian density describing the interaction of the
isobar with the nucleon and the photon (the magnetic dipole transition)
 \cite{gasiorowicz-69}
\begin{equation}
L_{em}=e\frac{\mu}{m_N}F^{\nu\,\rho}(x)(\bar \psi_\nu(x)\gamma_{\rho}\gamma_5
\psi (x) + h.c. )\,,
\end{equation} 
where $e$ is the electron charge, $\alpha = e^2/4\pi = 1/137$, $m_N=0.94$
GeV is the nucleon mass, $F^{\nu\,\rho}(x)=\partial^\nu A^\rho (x) - \partial^\rho
A^\nu (x)$ is the electromagnetic field, $\psi_\nu(x)$ is the spinor-vector
isobar field and $\psi (x)$ is the spinor nucleon field.

The width of the radiative decay $\Delta\to\gamma N$ 
\begin{equation}
\Gamma (\Delta\to\gamma N,\,m_\Delta )= \alpha \left (\frac{\mu}{m_N}
\right )^2\omega^3(m_\Delta )\left (1+\frac{1}{3}\left (\frac{m_N}{m_\Delta}
\right )^2\right )\,,
\end{equation} 
where $\omega (m_\Delta)=m_\Delta\left (1-m_N^2/m_\Delta^2\right )/2=0.257$
GeV is the photon energy, $m_\Delta = 1.232$ GeV.

Using experimental data \cite{pdg-96} $\Gamma (\Delta\to\gamma N,\,
m_\Delta)=BR(\Delta\to\gamma N,\,m_\Delta)\cdot\Gamma_\Delta (m_\Delta)=0.58
\cdot 10^{-2}\cdot 0.12$ GeV $=0.7\cdot 10^{-3}$ GeV, one gets
$\mu^2/m_N^2=4.7$ GeV$^{-2}$.

Now one can calculate the amplitudes of the isobar electroproduction
($e^-N\to e^-\Delta$), see Fig. 1. We can conveniently use the helicity
amplitudes in the reaction center mass system
$A_{\lambda_\Delta\,\lambda_N}^{\lambda_e^\prime\,\lambda_e}$, where
$\lambda_N\,,\,\lambda_\Delta\,,\,\lambda_e$ and $\lambda_e^\prime$ are the
helicities of the nucleon, of the isobar, of the initial and final electrons
respectively. The isobar production amplitudes by the positron differ 
from the corresponding electroproduction amplitudes by sign only.

Let us write out the amplitudes essential to our consideration
\begin{eqnarray}
&& A_{\ \;\frac{3}{2}\,\ \;\frac{1}{2}}^{-\frac{1}{2}\,-\frac{1}{2}}=
A_{-\frac{3}{2}\,-\frac{1}{2}}^{\ \;\frac{1}{2}\,\ \;\frac{1}{2}}=
e^2\frac{\mu}{m_N}\sqrt{2(t-t_{min}(m))}\left (s-m_N^2\right )\frac{f(t)}{t}
\,,\nonumber \\[1pc]
&& A_{\frac{3}{2}\,\frac{1}{2}}^{\frac{1}{2}\,\frac{1}{2}}=
A_{-\frac{3}{2}\,-\frac{1}{2}}^{-\frac{1}{2}\,-\frac{1}{2}}=
e^2\frac{\mu}{m_N}\sqrt{2(t-t_{min}(m))}\left (s-m_N^2\right )\left (1-
\frac{m^2-m_N^2}{s-m_N^2}\right )\frac{f(t)}{t}\,,
\end{eqnarray} 
and
\begin{eqnarray}
&& A_{\ \;\frac{1}{2}\,-\frac{1}{2}}^{-\frac{1}{2}\,-\frac{1}{2}}=
A_{-\frac{1}{2}\,\ \;\frac{1}{2}}^{\ \;\frac{1}{2}\,\ \;\frac{1}{2}}=
e^2\frac{\mu}{m}\sqrt{\frac{2}{3}(t-t_{min}(m))}\ \left (s-m_N^2\right )
\frac{f(t)}{t}\,,
\nonumber \\[1pc]
&& A_{\frac{1}{2}\,-\frac{1}{2}}^{\frac{1}{2}\,\ \;\frac{1}{2}}=
A_{-\frac{1}{2}\,\ \;\frac{1}{2}}^{-\frac{1}{2}\,-\frac{1}{2}}=
e^2\frac{\mu}{m}\sqrt{\frac{2}{3}(t-t_{min}(m))}\ \left (s-m_N^2\right )\left
(1-\frac{m^2-m_N^2}{s-m_N^2}\right )\frac{f(t)}{t}\,,
\end{eqnarray} 
where $m$ is the isobar mass (an invariant mass of $\pi N$ or $\gamma N$ in
which the isobar decays), $t=-(k-k^\prime)^2=-(p^\prime -p)^2,\ s=(k+p)^2=
(k^\prime +p^\prime)^2$, $p\,,\,p^\prime\,,\,k$ and $k^\prime$ are the
nucleon, isobar, initial and final electron (positron) four-momenta
respectively, see Fig. 1, $f(t)=1/(1+2t)^2$ is the "dipole" formfactor of
the electromagnetic transition $N\to\Delta$, see, for example,
\cite{omelaenko-79,aznaurian-86}, and references quoted there. Hereafter
$t$ comes in units of GeV$^2$.

We ignore the amplitudes $A_{\frac{1}{2}\,\frac{1}{2}}^{\lambda\lambda}$,
$A_{-\frac{1}{2}\,-\frac{1}{2}}^{\ \>\,\lambda\ \>\,\lambda}$,
$A_{\frac{3}{2}\,-\frac{1}{2}}^{\lambda\ \>\,\lambda}$ and
$A_{-\frac{3}{2}\,\frac{1}{2}}^{\ \>\,\lambda\lambda}$,
which are proportional to $t$, and contributions in the amplitudes of Eqs. 
(3) and (4), which are also proportional to $t$. The magnitude of all omitted
contributions in the total cross section has the order of $1\% $\,.

The differential in $m$ and $t$ cross section of the process
$e^\mp N\to e^\mp\Delta\to e^\mp\pi N$
\begin{eqnarray}
&&\frac{d^2\sigma}{dmdt}=4\alpha^2\left (\frac{\mu}{m_N}\right )^2\frac{t-
t_{min}(m)}{t^2}(f(t))^2\times \nonumber \\[1pc]
&&\times\left (1+\frac{m_N^2}{3m^2}\right )\left [1-\frac{m^2-m_N^2}{s-m_N^2}
 + \frac{\left (m^2-m_N^2\right )^2}{2\left (s-m_N^2\right )^2}\right ]
\frac{m^2\Gamma (\Delta\to \pi N,\,m)}{\left |D_\Delta (m)\right |^2}\,,
\end{eqnarray} 
where the isobar propagator and the mass dependent isobar width have the
forms
\begin{eqnarray}
&& D_\Delta (m)=m^2-m_\Delta^2+im\Gamma (\Delta\to \pi N,\,m)\,,\nonumber
\\[1pc]
&&\Gamma (\Delta\to \pi N,\,m)=\Gamma (\Delta\to \pi N,\,m_\Delta)
\frac{m_\Delta}{m}\left (\frac{1+m_N/m}{1+m_N/m_\Delta}\right )^2\frac{2\left
(q(m)/q(m_\Delta)\right )^3}{1+\left (q(m)/q(m_\Delta)\right )^2}\,,
\nonumber \\[1pc]
&& q(m)=\frac{1}{2m}\sqrt{\left (m^2-\left (m_N+m_\pi\right )^2\right )
\left (m^2-\left (m_N-m_\pi\right )^2\right )}\ ,
\end{eqnarray} 
$m_\pi=0.14$ GeV is the pion mass, $q\left (m_\Delta \right )=0.225$ GeV. In
Eq. (6) we put $\Gamma_\Delta (m)=\Gamma (\Delta\to \pi N,\,m)$,
$\Gamma (\Delta\to \pi N,\,m_\Delta)=0.12$ GeV.
The $m$ distribution integrated over the interval  $t_{min}(m)\leq t\leq 
t_{max}(m)$ 
\begin{eqnarray}
&&\frac{d\sigma}{dm}=\sigma (m)=4\alpha^2\left (\frac{\mu}{m_N}\right )^2
\left (\ln\frac{t_{max}(m)\left (1+2t_{min}(m)\right )}{t_{min}(m)\left
(1+2t_{max}(m)\right )}-1+\frac{t_{min}(m)}{t_{max}(m)}-\right.\nonumber
\\[1pc]
&&\left.-\frac{11+24t_{min}(m)+24t_{min}^{\ 2}(m)}{6\left (1+2t_{min}(m)
\right )^3}+\frac{11+24t_{max}(m)+24t_{max}^{\ 2}(m)}{6\left (1+2t_{max}(m)
\right )^3} \right)\times\nonumber \\[1pc]
&&\times\left (1+\frac{m_N^2}{3m^2}\right )\left [1-
\frac{m^2-m_N^2}{s-m_N^2} + \frac{\left (m^2-m_N^2\right )^2}{2\left (s-
m_N^2\right )^2}\right ]\frac{m^2\Gamma (\Delta\to \pi N,\,m)}
{\left |D_\Delta (m)\right |^2}\,,
\end{eqnarray} 
where
\begin{eqnarray}
&& t_{max}(m)=-2m_e+\frac{1}{2s}\left \{\left (s-m^2+m_e^2\right )\left (s-
m_N^2+m_e^2\right )+\right.\nonumber\\[1pc]
&&\left. + \sqrt{\left (s-\left (m-m_e\right )^2\right )\left (s-
\left (m+m_e\right )^2\right )\left (s-\left (m_N-m_e\right )^2\right )
\left (s-\left (m_N+m_e\right )^2\right )}\right \}\,,\nonumber \\
&& t_{min}(m)=-2m_e+\frac{1}{2s}\left \{\left (s-m^2+m_e^2\right )\left (s-
m_N^2+m_e^2\right )-\right. \nonumber\\[1pc]
&&\left. -\sqrt{\left (s-\left (m-m_e\right )^2\right )\left (s-
\left (m+m_e\right )^2\right )\left (s-\left (m_N-m_e\right )^2\right )
\left (s-\left (m_N+m_e\right )^2\right )}\right \}\,,
\end{eqnarray} 
$m_e=0.51\cdot 10^{-3}$ GeV is the electron mass.

At $s=1.842$ GeV$^2$ (the electron energy is equal to $m_\phi/2$)
the integrated over all interval $m_\pi + m_N \leq m\leq\sqrt{s}-m_e$ 
total cross section $\sigma=2.95$ $\mu$b.
\footnote{Note, that the cross section under discussion is equal to 7.74 
$\mu$b at the c-$\tau$-factories ($s=4.644$ GeV$^2$) and to 11.63 $\mu$b at 
the $b$-factories ($s=17.8$ GeV$^2$) .}

As large as this magnitude of the cross section is caused by the "large"
logarithm in Eq. (8):
$\ln \left (t_{max}(m_\Delta)/t_{min}(m_\Delta)\right )=13.11$, where
$t_{max}(m_\Delta)=0.169,\ t_{min}(m_\Delta)=1.29\cdot m_e^2=0.336\cdot
10^{-6}$. The isobar $\Delta (1.232)$ has no rival in this energy region.

The distributions in the pion energy $E_\pi$, in the nucleon energy $E_N$
and in the pion (nucleon) momentum $q$ have the forms in the isobar rest
system
\begin{eqnarray}
&&\frac{d\sigma}{dE_\pi}=\sigma \left (E_\pi\right )=\frac{m\left
(E_\pi\right )}{\sqrt{m_N^2+E_\pi^2-m_\pi^2}}\sigma \left (m(E_\pi)\right )\,,\nonumber
\\[1pc]
&&\frac{d\sigma}{dE_N}=\sigma \left (E_N\right )=\frac{m\left (E_N\right )}{
\sqrt{m_\pi^2+E_N^2-m_N^2}}\sigma \left (m(E_N)\right )\,,\nonumber\\[1pc]
&&\frac{d\sigma}{dq}=\sigma (q)=\frac{qm(q)}{
\sqrt{\left (m_N^2+q^2\right )\left (m_\pi^2+q^2\right )}}\sigma (m(q))\,,
\end{eqnarray} 
where $m\left (E_\pi\right )=E_\pi + \sqrt{m_N^2+E_\pi^2-m_\pi^2}$\,,
$m\left (E_N\right )=E_N + \sqrt{m_\pi^2+E_N^2-m_N^2}$\,,
$m(q)=\sqrt{m_\pi^2+q^2}+\sqrt{m_N^2+q^2}$\,,\, $q(m(q))=q$ .

The distribution $\sigma \left (E_\pi\right )$ is shown in Fig. 2.

The amplitudes from Eqs. (3) and (4) allow us to build up the isobar spin
density matrix and the angle distributions of the decay products in the
helicity system, that is in the isobar rest system with the quantization
direction along the isobar three-momentum in the reaction center mass system.

The integrated over azimuth angle distribution in the $\Delta\to\pi N$ decay
\begin{equation}
\frac{dW^{\pi (N)}}{d\cos \theta}= \frac{1}{1+\frac{m_N^2}{3m^2}}\left\{
\frac{3}{4}\left (1+\frac{m_N^2}{9m^2}\right )\sin^2\theta +
\frac{m_N^2}{3m^2}\cos^2\theta\right\}\,,
\end{equation} 
where $\theta$ is the angle between the pion three-momentum direction
(nucleon) and the quantization direction.

At $m=m_\Delta$
\begin{equation}
\frac{dW^{\pi (N)}}{d\cos \theta}= 0.67\sin^2\theta +
0.16\cos^2\theta\,.
\end{equation} 

Note, that the average of Eq. (10) over $m$ changes the coefficients in Eq. 
(11) less than by one percent.

In deriving Eq. (10), we used the phenomenological Lagrangian density
describing the interaction of the isobar with the nucleon and the pion
\cite{gasiorowicz-69}
\begin{equation}
L=\frac{G}{m_N}(\bar \psi_\nu(x)\psi (x)\partial^\nu\phi (x) + h.c. )\,,
\end{equation} 
where $\phi (x)$ is the pion field.

In the reaction center mass system, 64 \% of the isobars are  emitted
at an angle less than $10^\circ$  with reference to the beam axis, that is
why the quantization direction is close to the beam direction. If to take
into account that the isobar momentum is not large (the order of 100 MeV) then
it is clear that the pion (nucleon) emission largely across the quantization
direction, see Eq. (10), causes their emission largely across the beam
direction.

In the decay $\Delta^+ (1.232)$ ($\Delta^0 (1.232)$), there are produced the
$\pi^0$-mesons twice as large as the $\pi^+$ $(\pi^-)$-ones. This property
can be used to analyze the residual gas composition.

The integrated over azimuth  angle distribution in the $\Delta\to\gamma  N$
decay
\begin{eqnarray}
&&\frac{dW^{\gamma (N)}}{d\cos \vartheta}= \frac{1}{\left (1+
\frac{m_N^2}{3m^2}\right )^2}\left\{\frac{1}{4}\left (1+\frac{2m_N^2}{m^2}+
\frac{m_N^4}{9m^4}\right )\sin^2\vartheta +\left (1+\frac{m_N^4}{9m^4}\right )
\cos^2\vartheta\right\}=\nonumber\\[1pc]
&&= 0.386\sin^2\vartheta + 0.728\cos^2\vartheta\,,
\end{eqnarray} 
where $\vartheta$ is the angle between the photon (nucleon) three-momentum
direction and the quantization direction.

To get the energy and momentum distributions in the process 
$e^\mp N\to e^\mp\Delta\to e^\mp\gamma N$ it needs to substitute 
\begin{eqnarray}
&&\Gamma (\Delta\to \pi N,\,m)\to \Gamma (\Delta\to\gamma N,\,m)=\nonumber
 \\[1pc]
&&BR(\Delta\to\gamma N,\,m_\Delta)\cdot\Gamma_\Delta (m_\Delta)
\frac{m_\Delta}{m}\left (\frac{1+m_N^2/3m^2}{1+m_N^2/3m_\Delta^2}\right )
\frac{2\omega(m)^3/\omega (m_\Delta)^3}{1+\omega (m)^2/\omega (m_\Delta)^2}
\end{eqnarray} 
in Eqs. (5) and (7) and $E_\pi\to\omega\,,\,q\to\omega$ and $\,m_\pi\to 0$ in
Eq. (9), where $\omega$ is the photon energy in the isobar rest system, 
$\omega (m)=m\left (1-m_N^2/m^2\right )/2$.

The expected number of the produced $\Delta$-isobars per unit time and per 
unit length of a vacuum chamber
\begin{displaymath}
 N = {2I \over e} n \sigma \biggl[ {1 \over \mbox{m} \cdot \mbox{s}} \biggr],
 \end{displaymath}
where $n$ is a density of nucleons participating in the interaction in the 
vacuum chamber, $I$ is the electric current in a single beam.

The nucleon density $n$ is determined by the pressure of the residual gas and
by its partial composition.
The residual gas pressure $p \simeq 3\cdot 10^{-9} \mbox{Torr}$ \cite{vepp5}.
The partial composition of the residual gas, typically, is 50\% of 
$\mbox{H}_2$, 30\% of $\mbox{CO}$ and 20\% of $\mbox{CO}_2$ \cite{vepp5}.
The molecule density can be evaluated by the formula
\begin{displaymath}
 p=n_{M}kT,
\end{displaymath}
where k is the Boltzmann constant, $T=300 \mbox{K}$ is the residual gas
temperature which is determinated by the temperature of the accelerator walls.
 Then one gets
 $n_{M} \simeq 10^{14} \biggl[ {1 \over \mbox{m}^3} \biggr]$.
An effective number of the nucleus nucleons, with which the electron 
(positron) interacts in this process, is equal to $A^{2/3}$, where $A$ is the 
total nucleon number in a nucleus. With consideration for the partial 
composition of the residual gas the effective density of the nucleons 
$n=7\cdot 10^{14}\biggl[ {\mbox{nucleon}\over \mbox{m}^3} \biggr]$.

So, for the current $I=1\mbox{A}$ and the cross section $\sigma = 3 
\mu\mbox{b}$, the number of the isobars produced per unit time and per unit
length  
\begin{displaymath}
 N = 2 \biggl[ { \mbox{events} \over {\mbox{m} \cdot \mbox{s} } } \biggr].
\end{displaymath}

Although this frequency of the counting is small in comparison with the
frequency of the counting from the $\phi$-meson resonance $\sim 1 \mbox{kHz}$, 
nevertheless, the events of this process can make the extraction of the rare
decays at the $\phi$-factories difficult.

At present the study of the process of the $\Delta$-isobar electroproduction
is beginning in the experiments with the SND detector at the accelerator
complex VEPP-2M in Novosibirsk.

This work was partly supported by grants RFBR, 94-02-05 188,
96-02-00 548, and INTAS-94-3986.

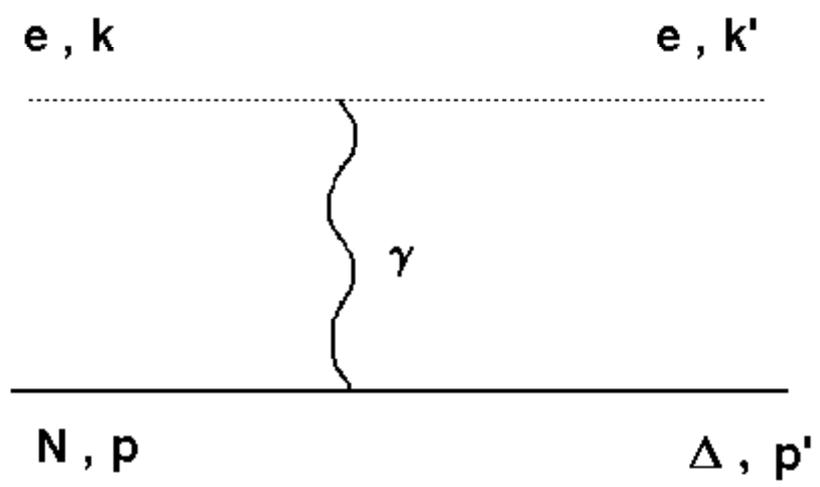
\begin{figure}
\caption{The diagram describing the isobar electroproduction.}
\end{figure}
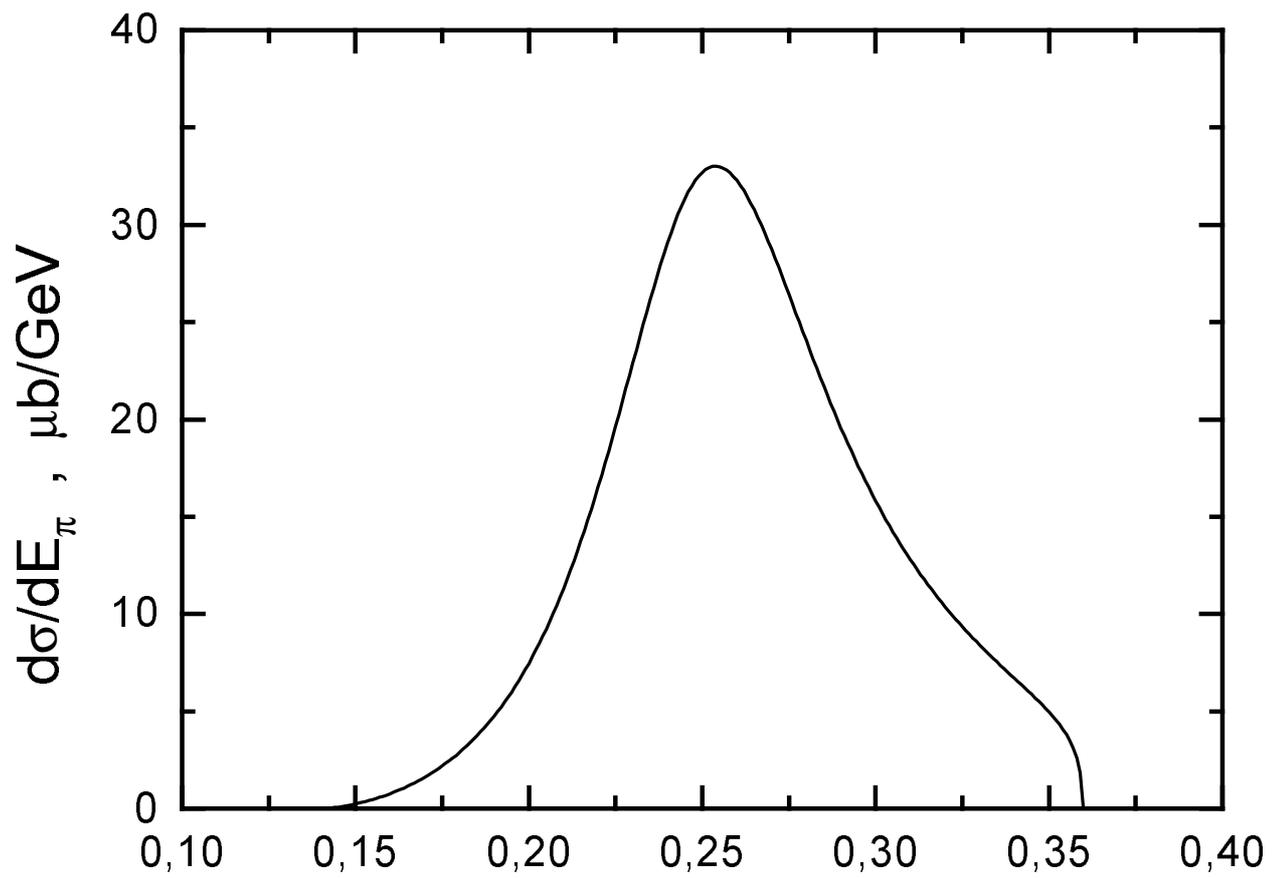
\begin{figure}
\caption{The pion energy distribution $\sigma \left (E_\pi\right )$ in the 
isobar rest system.}
\end{figure}

\end{document}